\documentclass{article}
\usepackage{cite}
\usepackage{amsmath,amssymb,amsfonts}
\usepackage{algorithm}
\usepackage{algorithmic}
\usepackage{booktabs}
\usepackage{color}
\usepackage{epstopdf}
\usepackage{epsfig}
\usepackage{float}
\usepackage{graphicx}
\usepackage{mathtools}
\usepackage{multicol}
\usepackage{multirow}
\usepackage{subfig}
\usepackage{tabulary}
\usepackage{textcomp}

\graphicspath{{figures/}}

\newcommand{\R}{\mathbb{R}}
\newcommand{\Rsq}{\mathbb{R}^2}
\newcommand{\Rcb}{\mathbb{R}^3}
\newcommand{\norm}[1]{||#1||}
\newcommand{\plos}{P_\textrm{LoS}}
\newcommand{\pnlos}{P_\textrm{NLoS}}
\newcommand{\mulos}{\mu_\textrm{LoS}}
\newcommand{\munlos}{\mu_\textrm{NLoS}}
\newcommand*{\QEDA}{\hfill\ensuremath{\blacksquare}}

\begin{document}

\title{Backhaul-Aware Optimization of UAV Base Station Location and Bandwidth Allocation for Profit Maximization}
\author{Cihan~Tugrul~Cicek, Hakan~Gultekin, Bulent~Tavli, Halim~Yanikomeroglu}

\maketitle

\begin{abstract}
Unmanned Aerial Vehicle Base Stations (UAVBSs) are envisioned to be an integral component of the next generation Wireless Communications Networks (WCNs) with a potential to create opportunities for enhancing the capacity of the network by dynamically moving the supply towards the demand while facilitating the services that cannot be provided via other means efficiently. A significant drawback of the state-of-the-art have been designing a WCN in which the service-oriented performance measures (e.g., throughput) are optimized without considering different relevant decisions such as determining the location and allocating the resources, jointly. In this study, we address the UAVBS location and bandwidth allocation problems together to optimize the total network profit. In particular, a Mixed-Integer Non-Linear Programming (MINLP) formulation is developed, in which the location of a single UAVBS and bandwidth allocations to users are jointly determined. The objective is to maximize the total profit without exceeding the backhaul and access capacities. The profit gained from a specific user is assumed to be a piecewise-linear function of the provided data rate level, where higher data rate levels would yield higher profit. Due to high complexity of the MINLP, we propose an efficient heuristic algorithm with lower computational complexity. We show that, when the UAVBS location is determined, the resource allocation problem can be reduced to a Multidimensional Binary Knapsack Problem (MBKP), which can be solved in pseudo-polynomial time. To exploit this structure, the optimal bandwidth allocations are determined by solving several MBKPs in a search algorithm. We test the performance of our algorithm with two heuristics and with the MINLP model solved by a commercial solver. Our numerical results show that the proposed algorithm outperforms the alternative solution approaches and would be a promising tool to improve the total network profit. \end{abstract}



\newpage
\section{Introduction}
\label{sec:introduction}

Unmanned Aerial Vehicle Base Stations (UAVBSs) are expected to be used in the next generation Wireless Communications Networks (WCNs) for enhancing the capacity of the network as well as expanding the coverage \cite{Andrews2014}. Although the rapid deployment and mobility advantage of a UAVBS has the potential to substantially improve the Quality-of-Service (QoS), there exist certain technical difficulties to be addressed such as resource management and channel modelling~\cite{Mozaffari2019}. Among others, the location of the UAVBSs play a key role to assist the WCN since positioning UAVBSs optimally has vital importance when compared to terrestrial base station positioning. In traditional terrestrial networks, base stations are typically located with respect to long-term traffic estimations. Even though these estimations vary in time, installing new cells should be a cost-effective choice to meet the demand. On the other hand, a UAVBS's location can be adjusted to meet instantaneous demand or to enhance the capacity of the WCN, yet, determining the optimal 3-D locations of such vehicle remains as a challenging and complex problem.

Not only does determining the locations of the UAVBSs improve the network performance, but also resource management should be jointly addressed to boost the benefit of UAVBSs. Therefore, there has been a number of attempts to jointly optimize the location and resource allocation decisions \cite{Azizi2019}. Indeed, it is important to take the finite backhaul capacity of the UAVBSs in consideration. Since the capacity is a significant concern for agility and reliability, joint optimization of the location and allocation decisions is imperative for achieving realistic assessment of the benefits of UAVBSs.

In this study, we, specifically, address the aforementioned drawbacks and consider a UAV-assisted WCN with a capacitated UAVBS serving to users on the ground who are not able to receive service from ground base stations (GBSs), e.g., due to high path loss. While serving the users, the backhaul capacity of the UAVBS and the available bandwidth of the WCN are also considered in the problem setup. The objective is to maximize the total system profit. In particular, the users are assumed to be offered different data rate levels to improve their satisfaction and each user is allowed to select at most one of the offered options. This approach can also be used as a new pricing strategy in the next generation WCNs since the extent of the offered service type is a significant factor to ensure both customer loyalty and satisfaction \cite{Mccarthy1964}. 

Moreover, many studies related to the UAVBS location have assumed that either the altitude or the projection of the utilized UAVBSs on the ground are fixed. Such approaches are shown to transform the problem into a lower computational complexity case. However, such simplifications would cause sub-optimal decisions. For instance, it is shown that allowing the UAVBSs to move both vertically and horizontally boosts the network performance in terms of throughput, resource utilization, and coverage \cite{Cicek2019b}. Therefore, an agile approach should be developed in which the assumptions that leads to sub-optimality are avoided to have more realistic results.

Although there have been some attempts to consider network profit as an objective in 5G \cite{Azizi2019}, this concept still needs to be thoroughly analyzed as the next generation WCNs are expected to embrace different pricing strategies. For instance, a non-cooperative pricing strategy can be adopted by the network operators when the resources are limited in a network \cite{Luong2019}. There exist some other strategies for pricing such as \cite{Luong2019,Gizelis2011} and the references therein, however, the aim of this paper is not to compare how different pricing strategies perform. Instead, we provide an optimization problem for a single network operator who compete in a selfish market and offer a range of data rate levels to maximize its profit based on a static service-level-based pricing strategy \cite{Gizelis2011}.

We summarize the contributions of this paper as follows:
\begin{itemize}
    \item \textbf{Backhaul-aware optimization:} Our proposed system considers both the backhaul and access capacities. Most of the studies related to the UAVBS location have studied uncapacitated UAVBS or assumed only one of these capacities, which would result in infeasible or unrealistic solutions. Therefore, we include both capacities in our problem setup so that the proposed model can be easily applied to real-life cases.
    \item \textbf{Profit maximization:} QoS-based optimization problems have covered a wide range of performance indicators such as throughput, latency or coverage \cite{Cicek2019a}. Although these indicators seem reasonable, there has been little incentive for the network operators to increase the willingness to use UAVBSs. We deliberately adopt the network profit as the objective function in the model so as to motivate network operators.
    \item \textbf{Efficient solution approach:} We use MINLP techniques to formulate the proposed model. Due to high complexity of this formulation, finding optimal solutions is difficult. We develop an efficient heuristic algorithm, which solves the problem within a reasonable time. We also provide the complexity of this algorithm.
    \item \textbf{Computational study:} We perform an extensive computational study in which we compare the performance of our algorithm with two more heuristics and one of the well-known MINLP solvers, the BARON solver over synthetically generated data. Our numerical results prove that our algorithm outperforms all three of the solution approaches.
    \item \textbf{Theoretical contributions:} We provide two important proofs regarding the concavity and unimodality of the data rate function, which is utilized extensively in the UAVBS literature. More specifically, (i) we prove that the data rate function is concave with respect to bandwidth and (ii) we prove that the data rate function is unimodal with respect to the UAVBS altitude. These results form the basis of proposed heuristic algorithm. 
\end{itemize}

The rest of the paper is organized as follows. Section~\ref{sec:liter} presents a literature overview on UAVBS location problems. The system model and mathematical formulation of the proposed system are given in Section~\ref{sec:sys-model}. In Section~\ref{sec:solution}, we propose the heuristic algorithm. Section~\ref{sec:results} presents the computational results and Section~\ref{sec:conclusion} concludes the paper with several future research directions. Proofs of concavity and unimodality of the data rate function are presented in Appendices~\ref{appxA}~and~\ref{appxB}, respectively.

\section{Literature review}\label{sec:liter}
The opportunities brought by the UAVBSs (such as enhancing capacity, improving QoS, extending coverage) are so promising that despite the relatively recent appearance of the topic, the literature on UAVBSs has grown rapidly. Especially, the location optimization problems have attracted significant interest since they have a significant impact on the network performance \cite{Chandrasekharan2016}. In this section, we present a high-level overview of the UAVBS location literature and present the differences of our study from the existing literature.

Recent studies on UAVBS location have considered several design challenges such as 3-D deployment, air-to-ground channel modelling, and trajectory optimization. For instance, \cite{Kalantari2017a} jointly minimizes the number of UAVBSs and determines the locations in areas with different user densities. In \cite{Bor-Yaliniz2016}, a mathematical model is proposed for the single UAVBS location problem, in which the ratio of the altitude of the UAVBS to the radius of the area is used to reduce the problem to a 2-D location problem. In \cite{Alzenad2017a}, the location of an uncapacitated UAVBS is determined to satisfy different QoS requirements of the users. In \cite{Mozaffari2016a}, the location of multiple UAVBSs is explored over a finite area to enhance the coverage and lifetime of the UAVBSs. In~\cite{ChenM2017}, a joint optimization problem in which the locations of the multiple UAVBSs and the association of the users to the UAVBSs are determined while the network is assumed to store caching information of the users. In~\cite{WuQ2018a}, the minimum average data rate provided to the users in a finite time horizon is maximized by jointly determining the trajectories of the UAVBSs and the user-UAVBS associations such that the maximum hovering distance is not violated. In fact, these studies, implicitly, assume that the backhaul capacity does not create a bottleneck when optimizing the UAVBSs which is justifiable in some cases.

While there exists a number of studies that investigate backhaul-aware UAVBS location optimization \cite{Kalantari2017a,Kalantari2017b,Kalantari2017c}, the network profit aspect of the UAVBSs have not been considered. In \cite{Kalantari2017b}, the location of a single UAVBS is considered to maximize the coverage such that the backhaul and access capacities are not exceeded. In \cite{Kalantari2017c}, the locations of multiple UAVBSs are determined to provide wireless service to the users who have a predefined delay-tolerance parameter. The resource allocation to the users in this study is also considered and an exhaustive search procedure is proposed to maximize a logarithmic utility function. However, in both studies, it is assumed that the users are identical in terms of QoS requirement. Since individual users are likely to have different requirement, a more realistic scenario would be assuming different demand values for users.

While the literature on UAVBS location is rich, there has been surprisingly little work on investigating how pricing schemes can boost the QoS in these networks. It is shown in \cite{Reichl2018} through comprehensive user trials on the quality perception that users are likely to pay for enhanced network quality. Hence, paying more for better service can be considered as a typical customer tendency.  Aligned with the strategy we propose in this study, it is shown that a well-designed pricing strategy, that does not depend only on flat-rate usage but also depends on the diversity and quality of services, can help network operators to improve profitability \cite{Soumya2013}. Pricing strategies can also help network operators improve the QoS. For instance, determining higher price levels for services that cause congestion could help network operators take control of excessive demand \cite{Bor-Yaliniz2019}.

Although the benefits of pricing strategy has been proven to be a significant factor in numerous studies, the related literature has been restricted to either uncapacitated network structures or a fixed capacity assumption. In the uncapacitated case, the UAVBSs are assumed to provide wireless services without considering the backhaul capacity or access capacity while the fixed capacity assumption only considers the bandwidth availability \cite{Azizi2019}. However, both of these capacities play significant roles for agile and reliable service provision. Therefore, we include both capacities in our system model.

The proposed model in this paper covers the three pillars of a well-designed WCN, which has not been jointly considered to date: the UAVBS deployment, backhaul and service-based pricing strategy. Since the state-of-the-art in UAV-assisted WCNs have mainly focused one or two of these pillars, the existing models and solution approaches cannot suit to a model in which all three pillars are combined. We, therefore, develop a novel UAV-assisted system model in the next sections and propose an efficient solution approach to the developed model.

\section{The System Model}
\label{sec:sys-model}
In our model, we consider a UAV-assisted WCN which consist of multiple GBSs and one UAVBS. The UAVBS has a dedicated backhaul link with one of the GBSs to serve the users on the ground. We assume that some of the users in the network cannot be served by GBSs due to several reasons (e.g., congestion in the wireless network or low channel quality). The UAVBS is used as a temporary service provider for those users. The users are offered different data rate options and allowed to select at most one option. The profit is earned based on the selection of a specific user. Each user can have a different willingness value to pay for the same service. This approach can be considered as a variation of ``Paris Metro Pricing'' (PMP) \cite{Luong2019}, in which the cumulative distribution function of willingness of the users is assumed to be known by the network operator instead of the exact willingness values. 

\begin{figure}[!t]
  \centering
  \includegraphics[width=\linewidth]{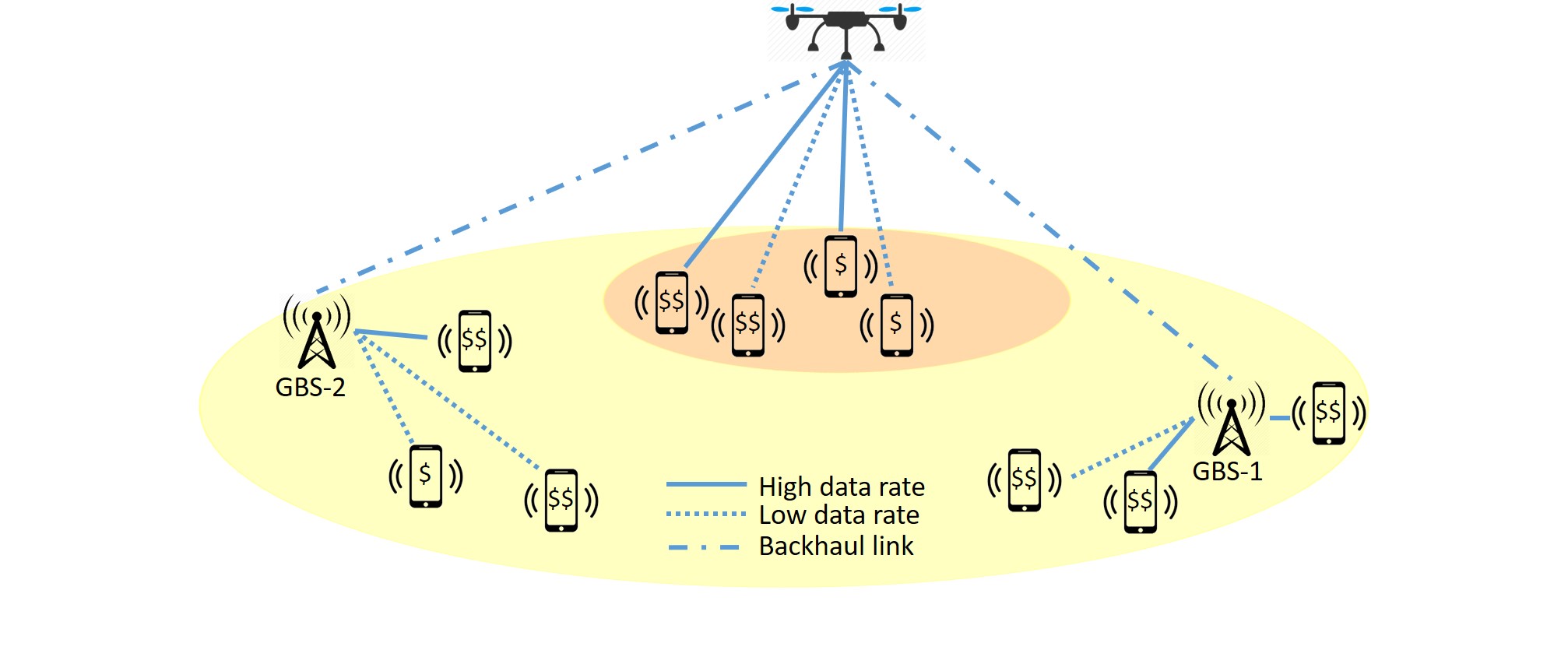}
  \caption{Illustration of the considered WCN architecture.}
  \label{fig:sample_network}
\end{figure}

Fig.~\ref{fig:sample_network} illustrates a sample representation of the considered system with two data rate options: low and high. The willingness values of the users to pay for the service are shown by the number of \$ icons in the figure. Although, the UAVBS can rapidly be deployed anywhere in the air to cover as many users as possible, the availability of a backhaul link should be considered carefully to provide a reliable connection. We assume that each GBS has sufficient connection capacity with the communications infrastructure to transmit the data received from the UAVBS to the core network. Therefore, there is no congestion in the fixed infrastructure.

The presented model can be validated based on the idea of enhancing the capacity as well as providing rapid supply to difficult-to-predict situations such as crowded events and activities. For instance, the area enclosed within the larger circle in Fig.~\ref{fig:sample_network} is a potential use case of such a model. The user distribution is not homogeneous, that is, some parts of the region hosts more users than the rest of the region. The UAVBS can serve users in the congested areas more efficiently than GBSs. The presence of a UAVBS also allows the network operators to utilize the idle capacity when the demand is not as high as the supply in some regions. Hence, the network operators can allocate the idle resources to the UAVBS to create a relay connection between the lightly loaded GBS and the users who cannot be served directly by that GBS.

In the rest of the paper, $I=\{1,\ldots,n\}$, $J=\{1,\ldots,m\}$ and $K=\{1,\ldots,s\}$ denote the set of users, GBSs, and offered data rate options, respectively. We assume that the service area of the UAVBS, $Q \subseteq \Rcb$, and the area in which users are located, $S\subseteq Q$, are finite. We use $X \in Q$ to denote the location of the UAVBS to be optimized. $y_i^u\in S$ and $y_j^g\in S$ denote the given locations of user $i$ and GBS $j$, respectively. Moreover, the available bandwidth of GBS $j$ that can be allocated to the UAVBS is assumed to be $b_j^g \geq 0$. 

The horizontal and vertical distances between two points, $x,y\in Q$, are found as $r=\norm{L(x-y)}$ and $h=\norm{M(x-y)}$, respectively, where $L$ and $M$ are linear transformations from $\Rcb$ to $\Rsq$ and $\R$, respectively, and $\norm{\cdot}$ is the $l_2$ norm. Throughout this paper, unless otherwise specified, we use boldface capital letters to denote matrices and lower-case letters to denote vectors consisting of scalar parameters or variables denoted by the same letter (e.g., $\mathbf{b^g} \in \R^m$ is the vector whose components are $b_j^g \geq 0$ for $j \in J$ and $\mathbf{Y^u} \in \R^{n \times 3}$ is the matrix whose components are $y_i^u \in S$ for $i \in I$). Moreover, the decision variables are denoted by upper-case letters, while the parameters are denoted with lower-case letters.

\subsection{Channel Model}
\label{sec:sys-model:channel}

There exist various air-to-ground channel models proposed in the literature \cite{Khawaja2019}. However, the most widely employed model is presented in \cite{Al-Hourani2014}, therefore, we adopt this model. According to this model, there exist two propagation regimes such that the users in the first regime can have Line-of-Sight (LoS) connections and the users in the second regime have Non-Line-of-Sight (NLoS) connections and can maintain their connections due to the mechanisms of electromagnetic wave propagation which can be used to convey information beyond the obstructions (e.g., reflection and diffraction). 

The probability of LoS is defined as a function of the locations of the UAVBS and the users (i.e., $\plos:Q\times S \rightarrow [0,1]$). This probability for a specific user located at $y^u\in S$ when the UAVBS is located at $X\in Q$ can be calculated as
\begin{equation}
    \plos(X,y^u) = \dfrac{1}{1 + {\alpha} e^{-\beta \left( \theta(X,y^u) - \alpha \right)}},     
\end{equation}

\noindent where $\alpha$ and $\beta$ are environment-specific constant parameters and $\theta(X,y^u) = (180\slash \pi)\arctan (h(X,y^u)\slash r(X,y^u))$ is the elevation angle. Then, the pathloss between a specific user located at $y^u\in S$ and the UAVBS located at $X\in Q$ is defined as
{\small
\begin{flalign} \label{eqn:pathloss-general}
    \nonumber L^u(X,y^u) & = 10\log_{10} \left( \dfrac{4\pi f_c d(X,y^u)}{c} \right)^\eta + \mulos\plos(X,y^u) \\ 
    \nonumber & + \munlos \pnlos \\
    \nonumber & = 10\log_{10} \left( \dfrac{4\pi f_c d(X,y)}{c} \right)^\eta + \mulos\plos(X,y^u) \\ 
    \nonumber & + \munlos (1-\plos(X,y^u)) \\
    & = A + 10\eta\log_{10} d(X,y^u) + B\plos(X,y^u),
\end{flalign}
}

\noindent where $d(X,y^u) = \norm{X-y^u}$ is the distance between $X\in Q$ and $y^u \in S$, $A = 10 \eta \log_{10} (4 \pi f_c \slash c) + \munlos$, and $B = \mulos - \munlos$ are constant parameters with $f_c$ denoting the carrier frequency in Hz, $c$ denoting the speed of light in m$\slash$s, $\eta$ denoting the path-loss component, $\mulos$ and $\munlos$ denoting the associated excessive pathloss in dB with probabilities $\plos$ and $\pnlos$, respectively, and $\pnlos = 1 - \plos$ \cite{Al-Hourani2014}. Note that the first term in the first equation in \eqref{eqn:pathloss-general} denotes the free-space pathloss.

Unlike the users, the GBSs are generally located at carefully determined and advantageous locations. This enables them to communicate with the UAVBSs through a much better channel. Therefore, the backhaul link is assumed to always have LoS communication with the GBSs, i.e., $\plos = 1$. Then, the pathloss between the UAVBS located at $X \in Q$ and a GBS located at $y^g\in S$ is calculated as
\begin{equation} \label{eqn:pathloss-GBS}
    L^g(X,y^g) = A + 10\eta \log_{10} d(X,y^g) + B.
\end{equation}

\subsection{Problem Formulation}
\label{sec:sys-model:formulation}
We consider point-to-point wireless connections between the UAVBS and the GBSs. This connection is assumed not to interfere with user links of the UAVBS. To facilitate such an assumption, reversed time-division duplexing is employed to avoid interference between the backhaul and user links, such that, during downlink of GBS, UAVBS is in the uplink mode. However, when GBS is in the uplink mode, users may largely be affected by the GBS due to the self-interference. To avoid this, orthogonal frequency channels are used in the backhaul and user links \cite{Wang2016}.

The objective in the proposed model is to maximize the network profit by jointly determining the UAVBS location and allocation of the available bandwidth to the users subject to the backhaul and access capacities. We first give the definitions of the data rate and backhaul capacity and then introduce the associated Mixed-Integer Non-Linear Programming (MINLP) formulation.

For a specific user located at $y^u\in S$, the actual data rate, when the allocated bandwidth is equal to $B^u\geq 0$ (in Hz) from the UAVBS located at $X \in Q$, can be found as
\begin{equation}
    R(X,y^u,B^u) = B^u \log_2 \left( 1 + 10^{S^u(X,y^u,B^u)}\right),
\end{equation}

\noindent where $S^u(X,y^u,B^u) =[ p^d - L^u(X,y^u) - 10\log_{10} B^u - \omega_\textrm{N}]\slash 10$ is the received signal-to-noise ratio (SNR) of user $i$ (in dB) with $p^d$ and $\omega_\textrm{N}$ denoting the transmit power of the UAVBS and a noise figure, respectively. Note that the data rate function, $R$, is concave with respect to $B^u$ (see Appendix~\ref{appxA} for the proof) and unimodal with respect to the altitude of the UAVBS (see Appendix~\ref{appxB} for the proof). This geometry will become a significant component of the solution approach we propose in Section~\ref{sec:solution}.

The backhaul capacity of a UAVBS located at $X \in Q$, when it has a backhaul link with a GBS located at $y^g \in S$, can be found as
\begin{equation}
    C(X,y^g) = b^g \log_2 \left( 1 + 10^{S^g(X,y^g)}\right),
\end{equation}

\noindent where $S^g(X,y^g) = [p^g - L^g(X,y^g) - 10\log_{10} b^g - \omega_\textrm{N}] \slash 10$ is the SNR of the UAVBS with $p^g$ denoting the transmit power of the GBS. Recall that the available bandwidth at GBSs, $\mathbf{b^g}$, is assumed to be known a priori.

Slightly abusing the notation, we use $R_i = R(X,y_i^u,B_i^u)$ and $C_j = C(X,y_j^g,b_j^g)$ to denote the actual data rate of user $i\in I$ and the backhaul capacity provided to UAVBS from GBS $j\in J$ when the UAVBS is located at $X \in Q$, respectively. We introduce the binary variable, $Z_j \in \{0,1\}$, to indicate whether or not the UAVBS has a backhaul link with GBS $j$, and the function, $u_i(R_i):\R \rightarrow \R$, to denote the profit gained when user $i \in I$ is served with data rate $R_i \geq 0$. Consequently, the MINLP formulation is defined as
\begin{flalign}
    \nonumber \mathcal{P}: & \underset{\substack{X \in Q, \mathbf{B^u}\in\R_+^n \\ \mathbf{Z}\in\{0,1\}^m}}{\max} \Pi(X,\mathbf{B^u}) = \sum_{i\in I} u_i(R_i) \\
    \nonumber & \textrm{subject to:} \\
    & \sum_{i\in I} R_i \leq \sum_{j\in J} C_j Z_j, \label{eqn:org:cons1}\\
    & \sum_{i\in I} B_i^u \leq \sum_{j\in J} b_j^g Z_j, \label{eqn:org:cons2}\\
    & \sum_{j \in J} Z_j = 1. \label{eqn:org:cons3}
\end{flalign}

The objective function in $\mathcal{P}$ maximizes the total network profit, constraints~\eqref{eqn:org:cons1} and \eqref{eqn:org:cons2} ensure that the backhaul and access capacities are not exceeded, respectively, while constraint~\eqref{eqn:org:cons3} requires the UAVBS to have a backhaul link with exactly one GBS.

\subsection{Pricing Model}
As explained in Section~\ref{sec:introduction}, we adopt a PMP-like pricing strategy, in which users are assumed to have different willingness values to pay for different data rate options. Therefore, the profit gained from user $i\in I$ is modeled as 
\begin{flalign} \label{eqn:cons-userrevenue}
    u_i(R_i) & =
    \begin{cases}
      0, & \hspace{-0.08in} R_i < \delta_1\\
      \phi_{ik}, & \hspace{-0.08in} \delta_k \leq R_i < \delta_{k+1}, k=1,\ldots ,s-1\\
      \phi_{is}, & \hspace{-0.08in} \delta_s \leq R_i \\
    \end{cases},
\end{flalign}

\noindent where $\delta_k$ denotes the $k^\textrm{th}$ data rate option in $K$, and $\phi_{ik}$ denotes the willingness value of user $i\in I$ to pay for the data rate option $k\in K$. The $\phi$ values are assumed to be an ascending order (i.e., $\phi_{i1}\leq \phi_{i2}\leq \ldots \leq \phi_{is}$), thus, providing higher data rate to a specific user would yield higher profit. This approach is based on the fact that the price of a service, generally, increases when its quality is improved \cite{Reichl2018}. Moreover, each user is assumed to have different willingness values for the same option, so that a diverse population can be represented.

$\mathcal{P}$ is a complex problem to solve since it includes a non-convex constraint set as well as binary and continuous variables. In fact, it belongs to the NP-complete problem class since relaxing the backhaul capacity constraint~\eqref{eqn:org:cons1} and fixing the bandwidth allocations would reduce the problem to the maximal covering location problem which is known to be NP-hard \cite{Berman2010}. Therefore, in the next section, we propose an iterative solution approach with lower complexity.

\section{Solution Approach}
\label{sec:solution}
In this section, we propose a heuristic algorithm in which $\mathcal{P}$ is solved in an iterative manner. Note that the backhaul capacity can be explicitly found when the UAVBS location is fixed, which substantially decreases the complexity of the problem. We exploit this relaxation and develop an efficient algorithm in which the altitude and the horizontal coordinates of the UAVBS are searched and at each fixed altitude several resource allocation problems are optimally solved for some fixed coordinates.

\subsection{Preliminaries}\label{subsec:solution:preliminaries}
Let $\overline{C}_j(\overline{X})$ be the actual backhaul capacity received from GBS $j$ given the UAVBS location, $\overline{X}\in Q$. We introduce the binary variable, $T_{ik} \in \{0,1\}$, to indicate whether or not user $i\in I$ is served with data rate option $k\in K$. There does not exist a closed form expression to determine the required bandwidth to serve a user located at $y^u\in S$ for a fixed data rate value, $\overline{\delta}$. However, since we proved that $R$ is concave with respect to $B^u$, this bandwidth can be explicitly found with a line search algorithm (e.g., bisection search). Let $\overline{B}_{ik}^u(\overline{X})$ denote the required bandwidth to serve user $i\in I$ with data rate option $k\in K$ when the UAVBS location is given as $\overline{X}\in Q$. Then, $\mathcal{P}$ can be reduced to the following Integer Linear Programming (ILP) formulation,
\begin{flalign}
    \nonumber \overline{\mathcal{P}}(\overline{X}) : & \underset{\substack{\mathbf{T}\in\{0,1\}^{n\times s} \\ \mathbf{Z} \in \{0,1\}^m}}{\max} \sum_{i\in I} \sum_{k\in K} \phi_{ik} T_{ik} \\
    \nonumber & \textrm{subject to:} \\
    & \sum_{i\in I} \sum_{k\in K} \delta_k T_{ik} \leq \sum_{j\in J}\overline{C}_j(\overline{X}) Z_j, \label{eqn:relax:cons1} \\
    & \sum_{i\in I} \sum_{k\in K} \overline{B}_{ik}^u(\overline{X}) T_{ik} \leq \sum_{j\in J}b_j^g Z_j, \label{eqn:relax:cons2} \\
    & \sum_{k\in K} T_{ik} \leq 1,\phantom{..} i \in I, \label{eqn:relax:cons3} \\
    & \sum_{j\in J} Z_j = 1. \label{eqn:relax:cons4}
\end{flalign}

The objective function in $\overline{\mathcal{P}}$ maximizes the total network profit for a fixed UAVBS location. Constraints~\eqref{eqn:relax:cons1} and \eqref{eqn:relax:cons2} ensure that the backhaul and access capacities are not exceeded. Constraints~\eqref{eqn:relax:cons3} stipulate that each user can only be served with at most one data rate option, while constraint~\eqref{eqn:relax:cons4} guarantees that the UAVBS has a backhaul link with exactly one GBS. 

The most important advantage of $\overline{\mathcal{P}}$ is that the formulation is now free of the non-convex data rate and backhaul capacity functions. In this way, instead of determining the actual data rate values of users with the location and allocation variables, the problem is altered to an assignment problem in which the data rate options are assigned to users regarding their profit values. Note that, $\overline{\mathcal{P}}$ can be solved separately for each GBS $j\in J$. The optimal solution is found when the UAVBS has a backhaul link with the GBS that provides the highest backhaul capacity, i.e., the highest value in the right-hand side in \eqref{eqn:relax:cons1}. This would allow the UAVBS to serve more users in case all other variables are fixed. Therefore, it is sufficient to solve this relaxed problem by setting the $Z$ variable corresponding to the GBS that yields the highest backhaul capacity for the given UAVBS location to 1 and all other $Z$ variables to 0.

Let $j^\ast(\overline{X})$ be the GBS providing the highest backhaul capacity for a given $\overline{X}$, i.e., $j^\ast(\overline{X}) = \arg \max_{j\in J} C(\overline{X},y_j^g,b_j^g)$. Then, omitting the $Z$ variables and associated constraint~\eqref{eqn:relax:cons4}, $\overline{\mathcal{P}}$ can be reformulated as
\begin{flalign}
    \nonumber \overline{\mathcal{P}}_{j^\ast(\overline{X})} : & \underset{\mathbf{T}\in\{0,1\}^{n\times s}}{\max} \overline{\Pi}(\mathbf{T}|\overline{X}) = \sum_{i\in I} \sum_{k\in K} \phi_{ik} T_{ik} \\
    \nonumber & \textrm{subject to:} \\
    & \nonumber \textrm{Constraints}~\eqref{eqn:relax:cons3}, \\
    & \sum_{i\in I} \sum_{k\in K} \delta_k T_{ik} \leq \overline{C}_{j^\ast(\overline{X})}, \label{eqn:relax:cons6} \\
    & \sum_{i\in I} \sum_{k\in K} \overline{B}_{ik}^u(\overline{X}) T_{ik} \leq b_{j^\ast(\overline{X})}^g. \label{eqn:relax:cons7}
\end{flalign}

The resulting $\overline{\mathcal{P}}_{j^\ast(\overline{X})}$ is also an ILP formulation, which can be efficiently solved by commercial solvers such as CPLEX or GUROBI. In fact, this problem is known as ``Multidimensional Binary Knapsack Problem'' (MBKP), which can be solved in pseudo-polynomial time via dynamic programming techniques \cite{Freville2004}. In MBKP, the problem is to select a subset of items subject to multiple capacity constraints. The objective is to maximize the total utility of the selected items that have specific utility values such that none of the capacity constraints are exceeded. Regarding our formulation in $\overline{\mathcal{P}}_{j^\ast(\overline{X})}$, items can be considered as user-data rate pairs, i.e., $T$ variables, while there exist two types of capacities, i.e., constraints~\eqref{eqn:relax:cons6} and \eqref{eqn:relax:cons7}. Therefore, items can be selected until either backhaul or access capacity is fulfilled. 

By exploiting the structure explained above and our proof of the unimodality of the data rate function with respect to the UAVBS altitude, we propose an iterative heuristic algorithm in the following subsection.

\subsection{Heuristic Algorithm}\label{subsec:solution:algorithm}

We proved that the data rate function is unimodal with respect to the UAVBS altitude. As a consequence of this fact, our heuristic algorithm uses a line search over the range of altitudes. At each altitude level, another search algorithm is used in which the revised relaxed formulation, $\overline{\mathcal{P}}$, is used to determine the optimal bandwidth allocations. The summary of the algorithm is given in Algorithm~\ref{algo:GSS}.

We utilize the well-known ``Golden Section Search'' (GSS) algorithm to search the optimal altitude of the UAVBS. Before explaining the details of the algorithm, we provide a brief description of the GSS. The GSS algorithm is typically used to find the global minimum or maximum of a unimodal function by iteratively searching the function domain. Suppose we are given a function, $f: \R \rightarrow \R$, that is assumed to have a single maximum within $[q,w]$, where $q,w\in\R$ and $-\infty < q < w < +\infty$. Then, two interior points are selected by using the golden ratio, $\gamma=\frac{\sqrt{5}-1}{2}$. The first point, $x_1$, is equal to $q+\gamma(q-w)$, and the second point, $x_2$, is equal to $w-\gamma(q-w)$. If $f(x_1) \geq f(x_2)$, then it is sufficient to say that the global maximum is between $x_2$ and $w$ and the same procedure continues within $[x_2,w]$, otherwise, the search continues within $[q,x_1]$. This recursive procedure continues until the range size drops below a predefined approximation threshold, $\epsilon_g$. As a result, the center of the final range is determined to be the maximum of $f$.

\begin{algorithm}[!t]
\caption{Finds the UAVBS location and the bandwidth allocations of the users}
\begin{algorithmic}[1]
	\renewcommand{\algorithmicrequire}{\textbf{Input:}}
	\REQUIRE $\mathbf{Y^u}$, $\mathbf{Y^g}$, $K$, $\mathbf{\phi}$, $\epsilon_g$, $h_l$, $h_u$.
    \WHILE {$h_u - h_l \geq \epsilon_g$} \label{algo:GSS:line1}
        \STATE $h_1 \leftarrow h_l + \gamma (h_u - h_l)$, $h_2 \leftarrow h_u - \gamma (h_u - h_l)$. \label{algo:GSS:line2}
        \STATE $(\overline{X}_1, \mathbf{\overline{B}^u_1}) \leftarrow \textrm{Algorithm}~\ref{algo:grid}(\mathbf{Y^u},\mathbf{Y^g},K, \mathbf{\phi},\mathcal{S}(h_1))$. \label{algo:GSS:line3}
        \STATE $(\overline{X}_2,\mathbf{\overline{B}^u_2}) \leftarrow \textrm{Algorithm}~\ref{algo:grid}(\mathbf{Y^u},\mathbf{Y^g},K, \mathbf{\phi},\mathcal{S}(h_2))$. \label{algo:GSS:line4}
        \IF{$\Pi(\overline{X}_1,\mathbf{\overline{B}^u_1}) \geq \Pi(\overline{X}_2,\mathbf{\overline{B}^u_2})$} \label{algo:GSS:line5}
            \STATE $h_l = h_2$. \label{algo:GSS:line6}
        \ELSE \label{algo:GSS:line7}
            \STATE $h_u = h_1$. \label{algo:GSS:line8}
        \ENDIF \label{algo:GSS:line9}
    \ENDWHILE \label{algo:GSS:line10}
    \STATE $h^\ast = (h_u + h_l)\slash 2$. \label{algo:GSS:line11}
    \STATE $(\overline{X}^\ast,\mathbf{\overline{B}^\ast}) \leftarrow \textrm{Algorithm}~\ref{algo:grid}(\mathbf{Y^u},\mathbf{Y^g},K, \mathbf{\phi},\mathcal{S}(h^\ast))$. \label{algo:GSS:line12}
    \RETURN $\overline{X}^\ast$, $\mathbf{\overline{B}^\ast}$.
\end{algorithmic}
\label{algo:GSS}
\end{algorithm}

\begin{algorithm}[!t]
\caption{Searches the horizontal plane for a fixed UAVBS altitude.}\label{algo:grid}
\begin{algorithmic}[1]
	\renewcommand{\algorithmicrequire}{\textbf{Input:}}
	\REQUIRE $\mathbf{Y^u}$, $\mathbf{Y^g}$, $K$, $\mathbf{\phi}$, $\mathcal{S}(\overline{h})$.
	\FOR{$\Tilde{s}_q \in \mathcal{S}(\overline{h})$} \label{algo:grid:line1}
	    \STATE $j^\ast = \arg \max_{j\in J}C(\Tilde{s}_q,y_j^g,b_j^g)$. \label{algo:grid:line2}
	    \STATE Solve $\overline{\mathcal{P}}_{j^\ast}(\Tilde{s}_q)$. Let $\mathbf{T^\ast_q}$ be the optimal solution. Set $\overline{\Omega}_q \leftarrow \overline{\Pi}(\mathbf{T^\ast_q}|\Tilde{s}_q)$. \label{algo:grid:line3}
    \ENDFOR \label{algo:grid:line4}
    \STATE $q^\ast \leftarrow \arg\max_{q=1,\ldots ,Q} \overline{\Omega}_q$. \label{algo:grid:line5}
    \RETURN $\Tilde{s}_{q^\ast}$, $\mathbf{\overline{B}^u}(q^\ast)$.
\end{algorithmic}
\end{algorithm}

Based on the above procedure, our algorithm determines two solutions at each iteration for the original problem by fixing the UAVBS altitude to two different levels. Let $h_l$ and $h_u$ denote the minimum and maximum altitudes to which the UAVBS can be located, respectively. The two altitude levels are determined as $h_1$ and $h_2$, i.e., $h_1 = h_l + \gamma(h_u-h_l)$ and $h_2 = h_u - \gamma (h_u - h_l)$. For each altitude level, a grid search algorithm, which is summarized in Algorithm~\ref{algo:grid}, is applied to determine the best UAVBS location and optimal bandwidth allocations at the corresponding altitude (Lines~\ref{algo:GSS:line3}-\ref{algo:GSS:line4} in Algorithm~\ref{algo:GSS}). Let $\mathbf{\overline{B}_1}$ and $\mathbf{\overline{B}_2}$, respectively, denote the optimal allocation decisions at $\overline{X}_1$ and $\overline{X}_2$, which are the best locations at altitudes $h_1$ and $h_2$, respectively. These locations and decisions are used to determine the direction of the GSS. If $\overline{X}_1$ yields a higher profit than $\overline{X}_2$ with the corresponding bandwidth allocations, then the altitude range is updated as $[h_2,h_u]$, otherwise, this range is updated to $[h_l,h_1]$. The GSS algorithm terminates whenever the size of this range drops below $\epsilon_g$. Note that as a last step, the relaxed problem is solved once more, where the UAVBS altitude is equal to the center of the final altitude range (Lines~\ref{algo:GSS:line11}-\ref{algo:GSS:line12} in Algorithm~\ref{algo:GSS}), and the best solution found at this altitude is reported.

In Algorithm~\ref{algo:grid}, for a fixed altitude level, $\overline{h}$, a grid set, $\mathcal{S}(\overline{h})$, is generated by dividing $S$ into $\mathcal{Q}$ smaller grids, i.e., $\mathcal{S}(\overline{h}) = \{\overline{s}_q \subseteq S:h = \overline{h}, q = 1,\ldots , \mathcal{Q}\}$, with $\Tilde{s}_q$ denoting the center of grid $q$. For each grid, the relaxed problem, $\overline{\mathcal{P}}$, is solved, where the UAVBS location is assumed as the center of the corresponding grid, i.e., $\overline{X} = \Tilde{s}_q$, $q=1,\ldots , \mathcal{Q}$. Then, the objective function values attained from all grids are compared and the center of the grid that yields the highest profit is returned as the best objective function value for $\overline{h}$. In this algorithm, $\overline{\Omega}_q$ denote the objective function value attained from grid $\overline{s}_q$. 

\subsection{Convergence and Computational Complexity}
The developed algorithm terminates after a finite number of iterations since the altitude range is assumed to be finite. Suppose that the initial altitude range is $[h_l^\textrm{init},h_u^\textrm{init}]$. Then, after each iteration, this range gets narrower by a constant rate, $\gamma \approx 0.618$, thus, after $k$ iterations, the altitude range is decreased to $(h_u^\textrm{init} - h_l^\textrm{init}) \times \gamma^k$. As $k$ goes to infinity, this range goes to 0. Since $\epsilon_g > 0$, the algorithm converges after a finite number of iterations.

From the computational perspective, we solve several MBKPs in each iteration, whose complexity depends on the number of items and the capacities \cite{Freville2004}. As we have $n\times s$ items for each problem and two capacities, $\overline{C}$ and $b^g$, the complexity of solving a single MBKP is $O(ns(\overline{C}+1)(b^g+1))$. The number of MBKPs to be solved for each fixed altitude is determined by $\mathcal{Q}$. Note that providing a dense grid structure would improve the solution accuracy, while it may substantially decrease the computational performance and vice versa. As a result, the overall complexity of our algorithm is $O(\mathcal{R}^- \mathcal{Q}ns(\overline{C}+1)(b^g+1))$, where $\mathcal{R}^-$ is the number of iterations to achieve $\epsilon_g$ approximation, i.e., $\mathcal{R}^- = \min\{\mathcal{R}\in \mathbb{Z}_+ : (h_u^\textrm{init} - h_l^\textrm{init})\gamma^\mathcal{R}\leq \epsilon_g\}$, where $\mathbb{Z}_+$ denotes the non-negative integers. 

\section{Computational Results}\label{sec:results}

In this section, a detailed computational study is performed to evaluate the performance of the developed algorithm in terms of solution accuracy and computational efficiency. We also provide some powerful insights on how service-based pricing can boost the network profit. All simulations are performed for a suburban area of 1500 m$\times$1500 m on an Intel i7-6700 CPU @3.40 GHz, 64-bit, 8GB RAM Windows 10 computer, where the communication and other simulation parameters are given in Table~\ref{table:parameters}. 
\begin{table}[!h]
\caption{Simulation parameters for suburban environment.}
\label{table:parameters}
\centering
\begin{tabular}{l l}
\toprule
Parameters & Values \\
\midrule
$\eta$, $\alpha$, $\beta$ & 2.5, 4.88, 0.43 \\
$\mu_\textrm{LoS}$, $\mu_\textrm{NLoS}$, $\omega_\textrm{N}$ & 0.1 dB, 21 dB, 6 dB \\
$f_c$ & 2 GHz \\
$p^g$ & 46 dBm \\
$b_j^g,j\in J$ & 10 MHz \\
$p^d$ & 36 dBm \\
$\epsilon_g$ & 1 m \\
$m$ & 4 \\
$n$ & 50, 75, 100, 125, 150, 175, 200 \\
$\mathcal{Q}$ & 50 \\
\bottomrule
\end{tabular}
\end{table}

For comparison, we use the optimal solutions (or the best feasible) attained from the well-known MINLP solver, the BARON solver \cite{Tawarmalani2004}, provided on the NEOS platform \cite{Czyzyk1998} with 1-hour time limit and two other heuristic approaches. Both our algorithm and heuristic algorithms are coded in Python v3.6. 

For the first heuristic algorithm, namely ``$\textrm{Heuristic}_\textrm{R}$'', the UAVBS location is randomly determined in $Q$ and optimal bandwidth allocations are determined by solving $\overline{\mathcal{P}}$ at this random location. To smooth the randomization effect, we use the best objective function values attained after 50 replications. For the second heuristic algorithm, namely ``$\textrm{Heuristic}_\textrm{F}$'', instead of a random location, the UAVBS is located to a fixed point, where the horizontal coordinates are determined as the weighted average coordinates of user locations and the altitude is determined as the quarters of the altitude range yielding four different altitudes. Consequently, the optimal bandwidth allocations are determined by solving $\overline{\mathcal{P}}$ at these four fixed locations and the location yielding the highest profit value is selected as the UAVBS location.

We assume that the GBSs are uniformly located in $S$ and the UAVBS altitude range is between 50 and 500 meters. The users are assumed to be located according to a Poisson Point Process (PPP). To apply PPP, the users are randomly located to specific parts of the region on $S$ with a random number of parent nodes and clustering rate. For instance, if the number of parent nodes is 5 and the clustering rate is 70\% for an instance with 100 users, then 70 users are located in close proximity of uniformly selected 5 different points while the remaining 30 users are located uniformly in $S$. 
The offered data rate options are given in Table~\ref{table:datarates}. The willingness values of the users are determined in a way that higher data rate values yield higher profit, i.e., $\phi_{ik} = \phi_{ik-1}+(\delta_k - \delta_{k-1}) \times \mathcal{U}(0,1),  k = 2, \ldots, s$ with $\phi_{i1} = \delta_1 \times \mathcal{U}(0,1)$, where $\mathcal{U}$ is the uniform random distribution. 
\begin{table}[!h]
\caption{Data rate options.}
\label{table:datarates}
\centering
\begin{tabular}{l l}
\toprule
No & Data rate values (Mbps) \\
\midrule
1 & \{1, 2\} \\
2 & \{1, 2, 4\} \\
3 & \{1, 2, 4, 8\} \\
\bottomrule
\end{tabular}
\end{table}

\begin{figure*}[!b]
    \centering
    \begin{tabular}{c c c}
        \subfloat[] {\includegraphics[width=.3\linewidth]{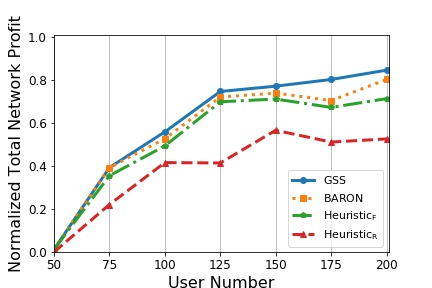}} &
        \subfloat[] {\includegraphics[width=.3\linewidth]{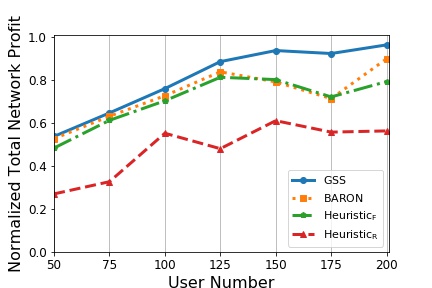}} &
        \subfloat[] {\includegraphics[width=.3\linewidth]{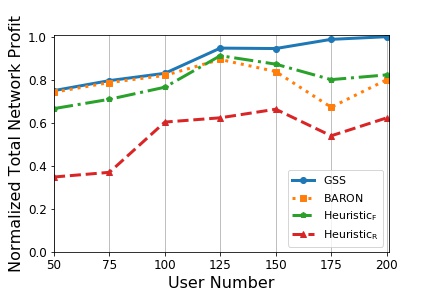}}
    \end{tabular}
    \caption{The normalized network profit with respect to different user numbers for data rate options (a) \{1, 2\} (b) \{1, 2, 4\} (c) \{1, 2, 4, 8\} (Mbps).}
    \label{fig:results1}
\end{figure*}

Since the locations of users and GBSs, and the willingness values are determined randomly, we simulate all instances with 10 replications, in which each instance refers to a single user number and data rate option set. Performances are reported based on the average of the replications for each instance. Hence, 210 different test problems are generated for 21 different instances, i.e., $n \in \{50,75,100,125,150,175,200\}$ and $|K| \in \{2,3,4\}$.

Fig.~\ref{fig:results1} depicts the change in the normalized objective function values for all solvers with respect to different user numbers for each data rate option set. The average CPU time of our algorithm is 255.1 seconds, while it decreases to 29.4 and 6.3 seconds for $\textrm{Heuristic}_\textrm{R}$ and $\textrm{Heuristic}_\textrm{F}$, respectively. The BARON solver reports the optimal solutions for only instances with 50 users and 2 data rate options, thus, we can only report the objective function values of the best feasible solutions whenever possible. 

Fig.~\ref{fig:results1} clearly shows that our algorithm outperforms all other solvers in terms of total profit. The $\textrm{Heuristic}_\textrm{R}$ is the worst performing algorithm because of the insufficient tolerance of avoiding local optima in determining the UAVBS location. In addition, the BARON solver suffers from increasing number of binary variables. Since the BARON solver uses a branch-and-bound technique over the binary variables, increasing number of these variables would significantly increase the number of branches and it would be inefficient to solve resulting non-convex optimization problem even after fixing some of the binary variable values on each branch. Note that our algorithm finds the optimal solutions for those instances for which the BARON also finds the optimum (2 instances and 20 replications). 

An interesting result observed from Fig.~\ref{fig:results1} is that the $\textrm{Heuristic}_\textrm{F}$ performance is better than that of $\textrm{Heuristic}_\textrm{R}$ although it is outperformed by our algorithm and BARON. The main drawback of $\textrm{Heuristic}_\textrm{F}$ is that it is not always the best alternative to locate the UAVBS to a central point, instead, optimally determining the location aligned with bandwidth allocations leads to a higher network profit. Nevertheless, the CPU times of $\textrm{Heuristic}_\textrm{F}$ is lower than the others which suggests that it can be employed to find a fast primal solution for large-scale problems. 

$\textrm{Heuristic}_\textrm{F}$ performance worsens with the number of data rate options. This is due to the trade-off between allocating more bandwidth to the users who yield higher profit in return and following a fairer allocation where the available bandwidth is allocated according to the central position of the UAVBS. Since the objective in our case is the network profit, the latter is advantageous. We have also observed that $\textrm{Heuristic}_\textrm{F}$ performance further worsens with higher clustering rates and parent node numbers since the weighted average location is unlikely to represent the overall dispersion of the users in such cases. Therefore, exploring the horizontal plane (e.g., grid search) appears to be a reasonable approach to improve the network profit. 

\begin{figure}[!h]
    \centering
    \includegraphics[width = .6\linewidth]{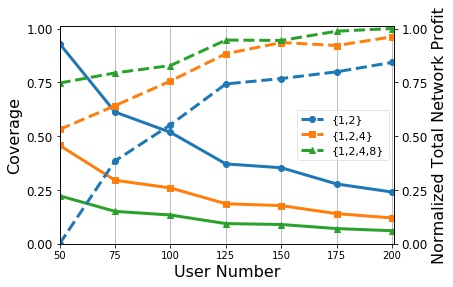}
    \caption{The change in the coverage and normalized total profit with respect to user numbers for different data rate options. The solid and dashed lines indicate the coverage and profit, respectively.}
    \label{fig:results2}
\end{figure}

In Fig.~\ref{fig:results2}, we present the change in coverage and total profit attained with respect to different data rate options. Note that the coverage is defined as the ratio of the total data rate provided to covered users over the maximum total data rate (i.e., $n \times \delta_s$). Fig.~\ref{fig:results2} shows that there exists a significant trade-off between improving the coverage and increasing network profit. As the likelihood of having more users with higher willingness values increases with more users, our algorithm prefers allocating the available bandwidth to those users who bring more profit which, in turn, leads to lower coverage.

\begin{figure}[!h]
    \centering
    \includegraphics[width = .6\linewidth]{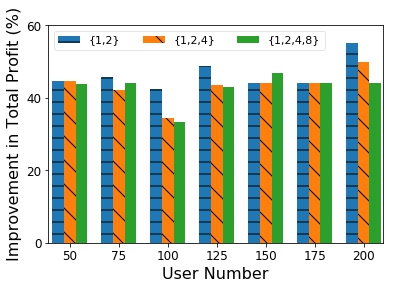}
    \caption{The improvement in the total profit with respect to user numbers for different data rate options.}
    \label{fig:results3}
\end{figure}

Fig.~\ref{fig:results3} verifies that designing a service-level-based pricing strategy improves the network profitability, substantially. Fig.~\ref{fig:results3} illustrates the change in the total profit between the solution found by our algorithm for the original problem in which more than one option is offered and a variant of the original problem where users are assumed to be offered a single data rate (that is equal to the average data rate offered in the original problem) and the profit gained from a specific user is assumed to be equal to the average willingness value of the corresponding user. The figure implies that the improvement level would increase with increasing user numbers. For instance, offering three data rate options to 50 users (e.g., 1, 2 and 4 Mbps) instead of offering a single option (e.g., 2.33 Mbps) improves the total profit by 44.6\%. This value escalates to 49.9\% when the user number increases to 200. Note that this outcome is valid for the assumption proposed in our system model that users are willing to pay more for improved service. Nevertheless, this assumption is highly likely to occur in real-life cases since a customer does not consider only the price but also does pay attention to the service quality \cite{Mccarthy1964,Reichl2018}.

\section{Conclusion}
\label{sec:conclusion}
UAVBS location problems have attracted significant interest in both industry and academia due to their potential to bring unprecedented advantages like rapid deployment, dynamic coverage extension, and on-demand capacity increase. Most of the studies to date have focused on improving QoS while the backhaul aspect and new pricing strategies are, mostly, overlooked. In this study, we address both the capacity aspect of QoS and a new pricing strategy based on different service levels. A novel mathematical model is developed and an efficient two-phase solution procedure, that combines GSS and grid search, is proposed for jointly optimizing the UAVBS location and bandwidth allocations to the users to maximize the network profit. The proposed algorithm is shown to be capable of significantly improving the network profit. 

This study can be extended by considering the temporal dimension of the problem since users typically relocate in time and the willingness values of the users, possibly, change at different time intervals. Such dynamics can lead to changes in the location, the hovering time, and the trajectory of the UAVBS which necessitates the design of dynamic pricing policies. Another extension can be to combine energy-related performance metrics with the backhaul capacity since energy-efficient operation of UAVBS is required to increase service time of the UAV-assisted WCNs. Covering all users with the minimum number of UAVBSs by considering new pricing strategies and backhaul capacity is also an interesting future research avenue.

\section*{Appendices}

\subsection{Proof of concavity of data rate function with respect to bandwidth}
\label{appxA}
In this appendix, we prove that the data rate function is concave with respect to $b_i$. By definition, a function is concave if the second derivative of the function is non-negative.   

Recall that the data rate function is defined as
\begin{flalign} 
   \nonumber R(X,y^u,B^u) & = B^u \log_2 \left( 1 + 10^{S^u(X,y^u,B^u)}\right),
\end{flalign}

\noindent where $S^u(X,y^u,B^u) =[ p^d - L^u(X,y^u) - 10\log_{10}B^u - \omega_\textrm{N}]\slash 10$. Since we will prove the concavity for only bandwidth, $B^u$, we can assume that the other variables are fixed, i.e., $X = \overline{X}, y^u = \overline{y}$. After a number of algebraic manipulations, we have
\begin{equation}\label{eqn:appx1}
    R(B^u) = B^u \log_2 \left(1+\frac{\Theta}{B^u}\right),
\end{equation}

\noindent where $\Theta = 10^{p_d - L^u(\overline{X},\overline{y}) - \omega_\textrm{N}-1} > 0$ is a fixed term. Having taken the second derivative of \eqref{eqn:appx1} with respect to $B^u$, we have
\begin{equation}
    R^{\prime\prime}(B^u) = -\frac{\Theta^2}{\log(2)B^u(B^u+\Theta)^2}.
\label{eqn:concavity}
\end{equation}

Since $B^u > 0$, $R^{\prime\prime}(B^u) < 0 $ always holds, the proof is complete. \QEDA

\subsection{Proof of unimodality of data rate function}
\label{appxB}

In this appendix, we prove that the data rate function is unimodal with respect to the UAVBS altitude. By definition, a function, $f:\R \rightarrow \R$, is unimodal if there exists an $m\in\R$ such that for some value, $\delta \in \R$, it is monotonically increasing (decreasing) for $\delta \leq m$ and monotonically decreasing (increasing) for $\delta \geq m$. In this case, the maximum (minimum) value of $f(\cdot)$ is $f(m)$ and there are no other local maxima (minima). If the function is differentiable, proving that the derivative of the function is equal to 0 at a single point in its domain proves its unimodality. Here we will make use of this property. 

Recall the data rate function is defined as
\begin{flalign} 
   \nonumber R(X,y^u,B^u) & = B^u \log_2 \left( 1 + 10^{S^u(X,y^u,B^u)}\right),
\end{flalign}

\noindent where $S^u(X,y^u,B^u) =[ p^d - L^u(X,y^u) - 10\log_{10}B^u - \omega_\textrm{N}]\slash 10$. Since we will prove the unimodality for only UAVBS altitude, $h$, we can assume that the other variables are fixed, i.e., $y^u = \overline{y}, B^u = \overline{B}, r(X,y^u) = \overline{r}$. After a number of algebraic manipulations, we have
\begin{flalign}
    R(h) & = \overline{B} \log_2 f(h),     
\label{eqn:appx2}
\end{flalign}

\noindent where $f(h) = 1 + \left( \overline{r}^2 + h^2\right)^{-\frac{\eta}{2}}m(h)$, $m(h) = KN^{g(h)} \geq 0$, $N=10^{\frac{\mu_\textrm{NLoS} - \mu_\text{LoS}}{10}} \geq 0$, $K = \overline{B}^{-1} \left(\frac{4\pi f_c}{c}\right)^{-\eta}10^{\frac{p^d-\mu_\text{NLoS}-\sigma_\textrm{N}}{10}} \geq 0$, $g(h)=\frac{1}{1+\alpha p s^{\theta(h)}} \in (0,1]$, $p=e^{\alpha \beta} \geq 0$, $s=e^{-\beta} \geq 0$, and $\theta(h)=\frac{180}{\pi}\arctan\left(\frac{h}{\overline{r}}\right) \in [0,90]$.

Having taken the first derivative of $R$ with respect to $h$, we have
\begin{flalign}
    R^\prime(h) & = \frac{\overline{B} f^\prime(h)}{\log(2)f(h)},    
\label{eqn:appx3}
\end{flalign}

\noindent where we denote a partial derivative with `` $^\prime$ '', e.g., $f^\prime(h)=\frac{\partial f(h)}{\partial h}$. Note that the denominator in \eqref{eqn:appx3} is non-negative, since $f(\cdot)$ is a product of two non-negative values plus 1. For a given bandwidth of $\overline{B} \geq 0$, this derivative can only be 0 if $f^\prime(h) = 0$. The first derivative of $f(h)$ can be found as
{\small\begin{flalign}
    f^\prime(h) & = \left(\overline{r}^2 + h^2\right)^{-\frac{\eta}{2}}\left[m^\prime(h) - \eta h \left(\overline{r}^2 + h^2\right)^{-1} m(h)\right].
\label{eqn:appx4}
\end{flalign}}

Since we assume $h > 0$, the first term in \eqref{eqn:appx4} is also greater than 0. Therefore, $f^\prime(h)=0$ holds if the expression in brackets is 0. Before proceeding, we first derive $m^\prime(h)$ as 
\begin{flalign}
    m^\prime(h)=K\log(N)N^{g(h)}g^\prime(h).
\label{eqn:appx5}
\end{flalign}

By substituting \eqref{eqn:appx5} into the expression in the brackets in \eqref{eqn:appx4}, we have 
\begin{flalign}
    \nonumber & K\log(N) N^{g(h)} g^\prime(h) - \eta h \left(\overline{r}^2 + h^2\right)^{-1} K N^{g(h_d)} \overset{?}{=} 0\\
    \Longleftrightarrow & \nonumber K N^{g(h)} \left(\log(N) g^\prime(h) - \eta h \left(\overline{r}^2 + h^2\right)^{-1}\right) \overset{?}{=} 0 \\
    \Longleftrightarrow & \log(N) g^\prime(h) \overset{?}{=} \eta h \left(\overline{r}^2 + h^2\right)^{-1}.\label{eqn:appx6}
\end{flalign}

The last transformation in \eqref{eqn:appx6} follows since both $K$ and $N$ are greater than 0. Before proceeding, we first derive $g^\prime(h)$ as
\begin{flalign}
    g^\prime(h) & = \frac{\alpha \beta p s^{\theta(h)} \theta^\prime(h)}{\left(1+\alpha p s^{\theta(h)}\right)^2},
\label{eqn:appx7}
\end{flalign}

\noindent where $\theta^\prime(h)=\frac{180\overline{r}}{\pi\left(\overline{r}^2 + h^2\right)}$. By substituting $\theta^\prime(h)$ into \eqref{eqn:appx7} and then \eqref{eqn:appx7} to \eqref{eqn:appx6}, we have
\begin{flalign}
    \nonumber & \frac{\log(N)180 \overline{r} \alpha \beta p s^{\theta(h)}}{\pi \left(1 + \alpha p s^{\theta (h)}\right)^2 \left(\overline{r}^2 + h^2\right)} \overset{?}{=} \frac{\eta h}{ \left(\overline{r}^2+h^2\right)} \\
    \nonumber \Longleftrightarrow & \frac{180(\mu_\textrm{NLoS} - \mu_\textrm{LoS}) \overline{r} \alpha \beta p s^{\theta(h)}}{\eta \pi \left(1 + \alpha p s^{\theta(h)}\right)^2} \overset{?}{=} h \\
    \nonumber \Longleftrightarrow & \frac{A s^{\theta(h)}}{\left(1 + \Lambda s^{\theta(h)}\right)^2} \overset{?}{=} h \\
    \nonumber \Longleftrightarrow & A q(h) = h \left(1 + 2 \Lambda q(h) + \left(\Lambda q(h)\right)^2\right) \\
    \Longleftrightarrow & q(h)(A - 2\Lambda h) \overset{?}{=} h \left(1 + \left(\Lambda q(h)\right)^2\right), \label{eqn:appx8}
\end{flalign}

\noindent where $A=\frac{180(\mu_\textrm{NLoS} - \mu_\textrm{LoS}) \overline{r} \alpha \beta p }{\eta \pi} \geq 0$, $\Lambda=\alpha p \geq 0$, and $q(h)=s^{\theta(h)} \in (0,1]$. For simplicity we use $Q(h)=q(h)(A-2\Lambda h)$ and $W(h)=h\left(1+\left(\Lambda q(h)\right)^2\right)$ to denote the left and right-hand side in \eqref{eqn:appx8}. Since $\theta(h)$ is a monotonically increasing function in $h$ and $\beta > 0$, $q(h) = s^{\theta(h)} = e^{-\beta\theta(h)}$ is a monotonically decreasing function in $h$. When $h$ starts to increase from 0, $Q(h)$ monotonically decreases from $A$ to 0. On the other hand, $W(h)$ is a monotonically increasing function which increases from 0 to the maximum altitude, $h^+$, when $h$ increases. This implies that there can be found at most one $h=\overline{h}$ where $Q\left(\overline{h}\right)=W\left(\overline{h}\right)$ which proves the unimodality of $R$ with respect to $h$. \QEDA

Also it is straightforward to show that $R(\cdot)$ monotonically increases for $h<\overline{h}$ and monotonically decreases for $h>\overline{h}$, which implies that $R(\cdot)$ has its maximum value when $h=\overline{h}$.


\bibliographystyle{ieeetr.bst}
\bibliography{access}

\end{document}